\documentclass[]{UCD_CS_FYP_Report}
\usepackage{graphicx}
\usepackage{listings}
\usepackage{hyperref}
\usepackage{amsmath}
\usepackage{multirow}
\usepackage{pdflscape}
\usepackage{booktabs}

\hypersetup{
    colorlinks=true,
    linkcolor=blue,
    filecolor=blue,      
    urlcolor=red,
    citecolor=blue,
}
\usepackage[backend=biber,style=alphabetic]{biblatex}
\def\projecttitle{Progressive Rock Music classification} 

\begin{document}
\maketitle

\tableofcontents
\pagenumbering{arabic}
\pdfbookmark[0]{Table of Contents}{toc}\newpage



\chapter{Introduction}
\pagenumbering{arabic}
\setcounter{page}{1}
\section{What is Prog Rock?}

Progressive rock, commonly referred to as \textit{prog} rock, is a genre of music that emerged in the late 1960s and early 1970s. \textit{Prog} rock is often characterized by its complex compositions, varied instrumentation, and influences from classical and jazz genres.

\textit{Prog} rock compositions are often highly intricate and sophisticated. They frequently incorporate complex time signatures, unconventional song structures, and elaborate arrangements. These songs often feature multiple sections with varying musical themes and motifs in contrast to the chorus-verse structure found in mainstream rock music. Time signatures such as 7/8 or even 5/4 are often used in contrast to 4/4 time frequented by other genres.

Instruments such as keyboards, synthesizers, flutes, and violins, are used in \textit{prog} rock in addition to traditional rock instruments like guitar, bass and drums.

\textit{Prog} rock also draws inspiration from a diverse range of musical genres, including classical music and jazz. Progressive rock often employs complex chord progression and harmonies reminiscent of classical music.

\section{Music Genre Classification}

Music genre classification is a critical task in the field of Music Information Retrieval (MIR), aimed at categorizing music into distinct genres based on audio features. This process is essential for various applications such as music recommendation, automatic playlist generation, and content-based music retrieval. \cite{10026394}

Traditional machine learning methods, such as Support Vector Machines and Random Forests, have been widely used in the problem of music genre classification. Deep learning methods, such as Convolution Neural Networks (CNNs) \cite{choi2017convolutional} and, more recently, Transformers \cite{abc}, have been used extensively to classify music. However, deep learning has emerged as the clear winner and will be the primary focus of our work.

This document is a report on our work to make machines differentiate between progressive rock music and other genres of music. Chapter \ref{Chap:2} explains how we clean and extract features from raw audio data. Chapter \ref{Chap:3} describes our Ensemble method experiments. Chapters \ref{Chap:4}-\ref{Chap:5} cover deep learning and discuss a 1D CNN model used to classify music. Chapter \ref{Chap:6} describes how we applied the Audio Spectrogram Transformer, a SOTA audio classification model. In Chapter \ref{Chap:7}, we present our findings and discuss them.

\chapter{Feature Extraction}
\label{Chap:2}
\section{Input Features}
 For the data collection and processing part, the songs were collected from directories containing both progressive rock and non-progressive rock songs. Each song was loaded using Librosa, which is a Python package for music and audio analysis. We use Librosa to load each song as well as pre-process it. For example, we trim silence from the audio, normalize it, and break the audio into smaller segments for analysis.

In regards to our feature extraction, we determined common attributes from our segments of audio. Spectrograms, MFCCs, chromograms, and beat position. Spectrograms are a visual representation of the frequencies of a signal as they vary with time. It provides insights into the frequency content of the audio signal over time. MFCCs are coefficients that capture the short-term power spectrum of a sound and are widely used in speech and audio processing tasks. Chromagrams represent the energy distribution of pitch classes in a given audio segment, providing information about the harmonic content of the audio. Beat position gives information which identifies the timing of beats within the audio. Beat positions are crucial for understanding the rhythmic structure of the music.

Next, we needed to prepare our model for input. Once the features are extracted for each audio segment, we combine them into a single tensor. This tensor represents the input for our model. Each row of our tensor corresponds to a segment of the song while each column represents a feature such as spectrogram, MFCCs, chromogram, or beat position. 

Now we must discuss the data splitting portion of the project as well as representing our data in a chart form. We split the dataset into training and validation sets so we may use one dataset to train our model and the other to test our model’s accuracy. 20\% of the songs from our dataset are randomly selected to be present in the validation set while the other 80\% are used to train the model. Having this diverse split of our data ensures our model is trained and evaluated on a slew of examples, which can help us assess it generalization capability. After we split the data, our script counts the number of songs in both our training and validation sets. This information is then visualized using a chart, which helps understand the balance between the training and validation data. Using a chart helps us visually verify that both the training and validation sets are adequately representing the entire dataset. 

\newpage

\begin{figure}
    \centering
    \includegraphics[width=1\linewidth]{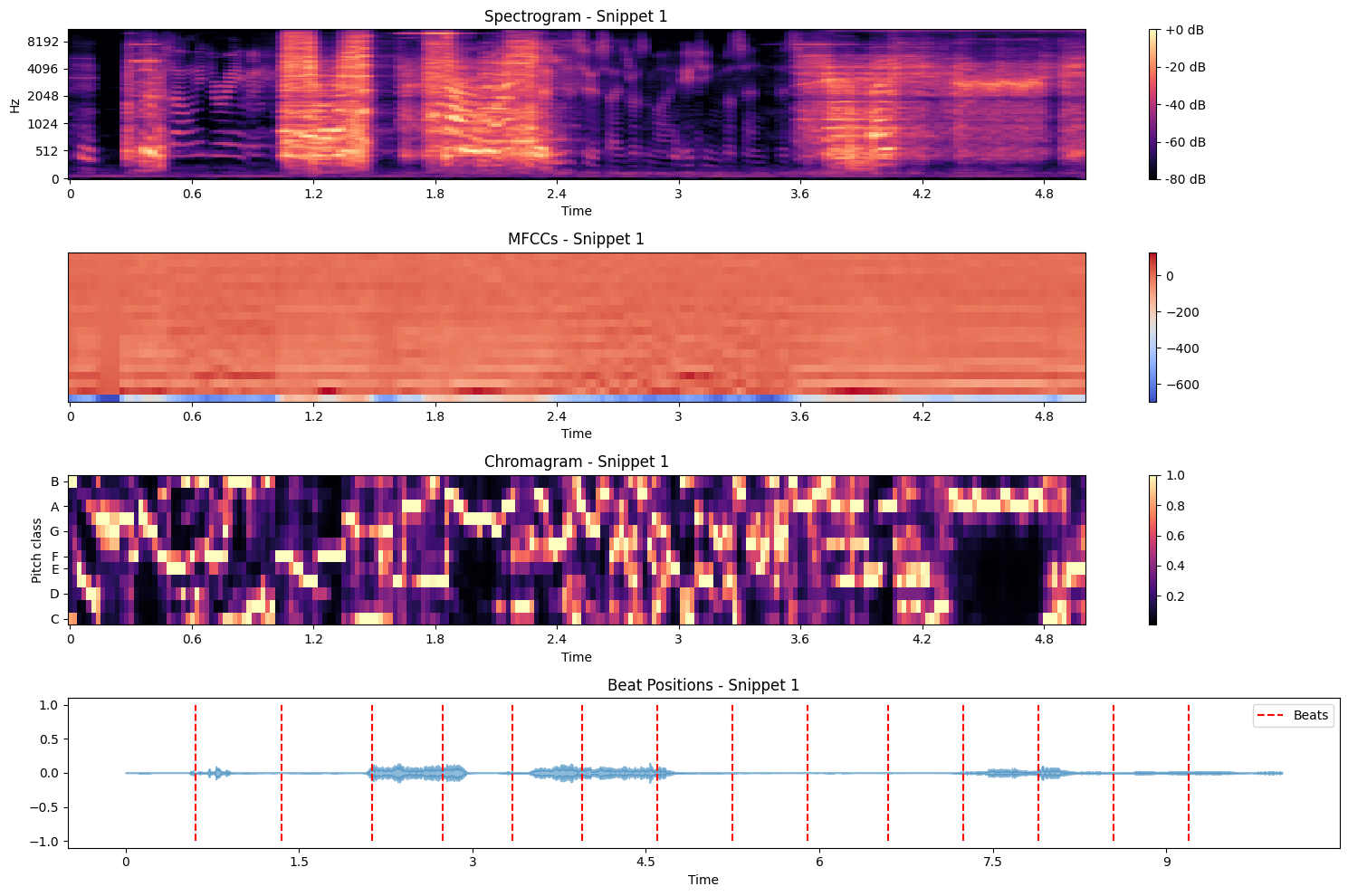}
    \caption{Visualizing sequence input (i) Spectrogram (ii) MFCCs (iii) Chromagram (iv) Beat Position for a 10 second snippet of Another One Bites the Dust by Queen}
    \label{fig:enter-label}
\end{figure}

\newpage

\section{Winner-take-all voting strategy}

\begin{figure}
    \centering
    \includegraphics[width=1
\linewidth]{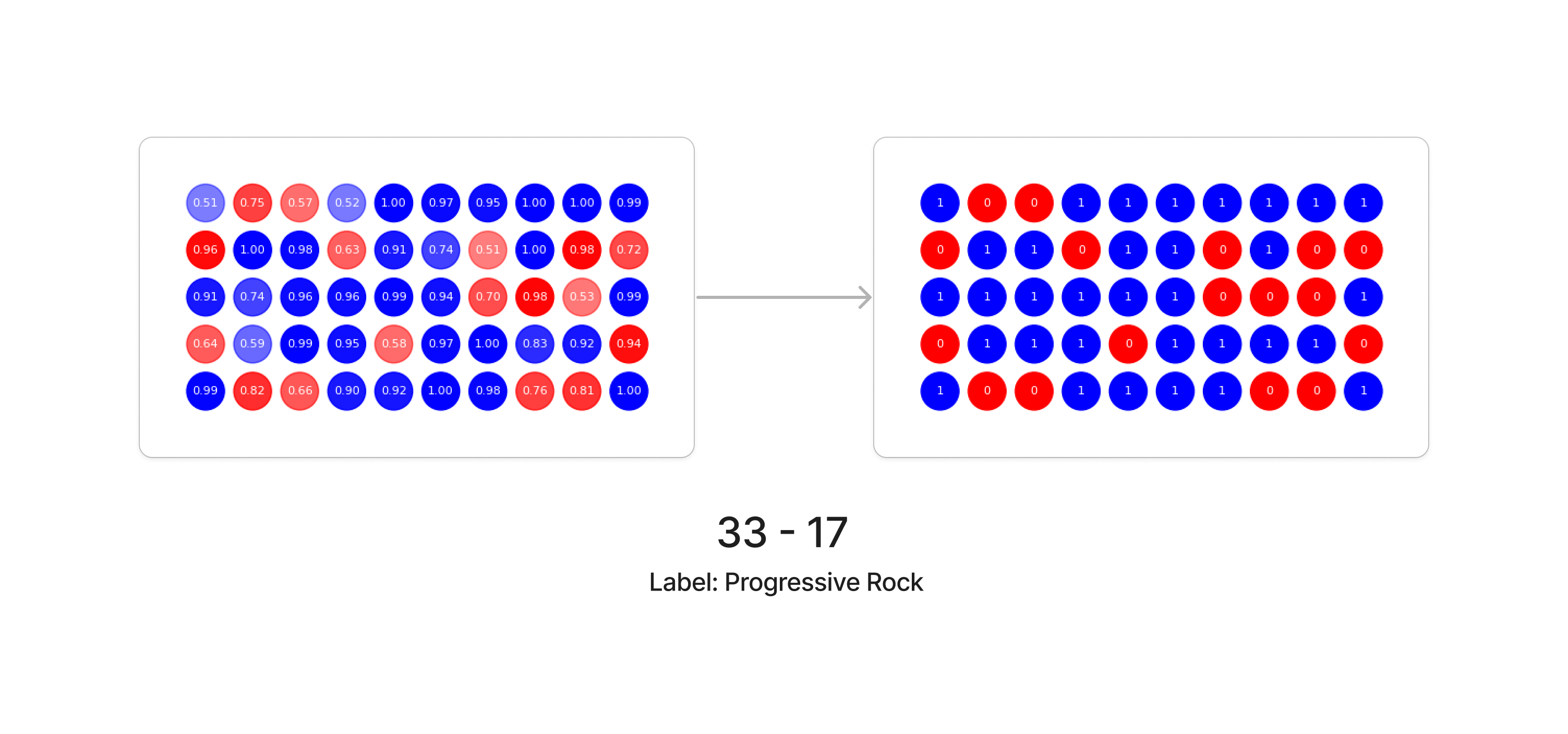}
    \caption{Classifying 50 Snippets of Toxicological Whispering by Amon Düül II using our winner-take-all voting strategy}
    \label{progsong}
\end{figure}

\begin{figure}
    \centering
    \includegraphics[width=\linewidth]{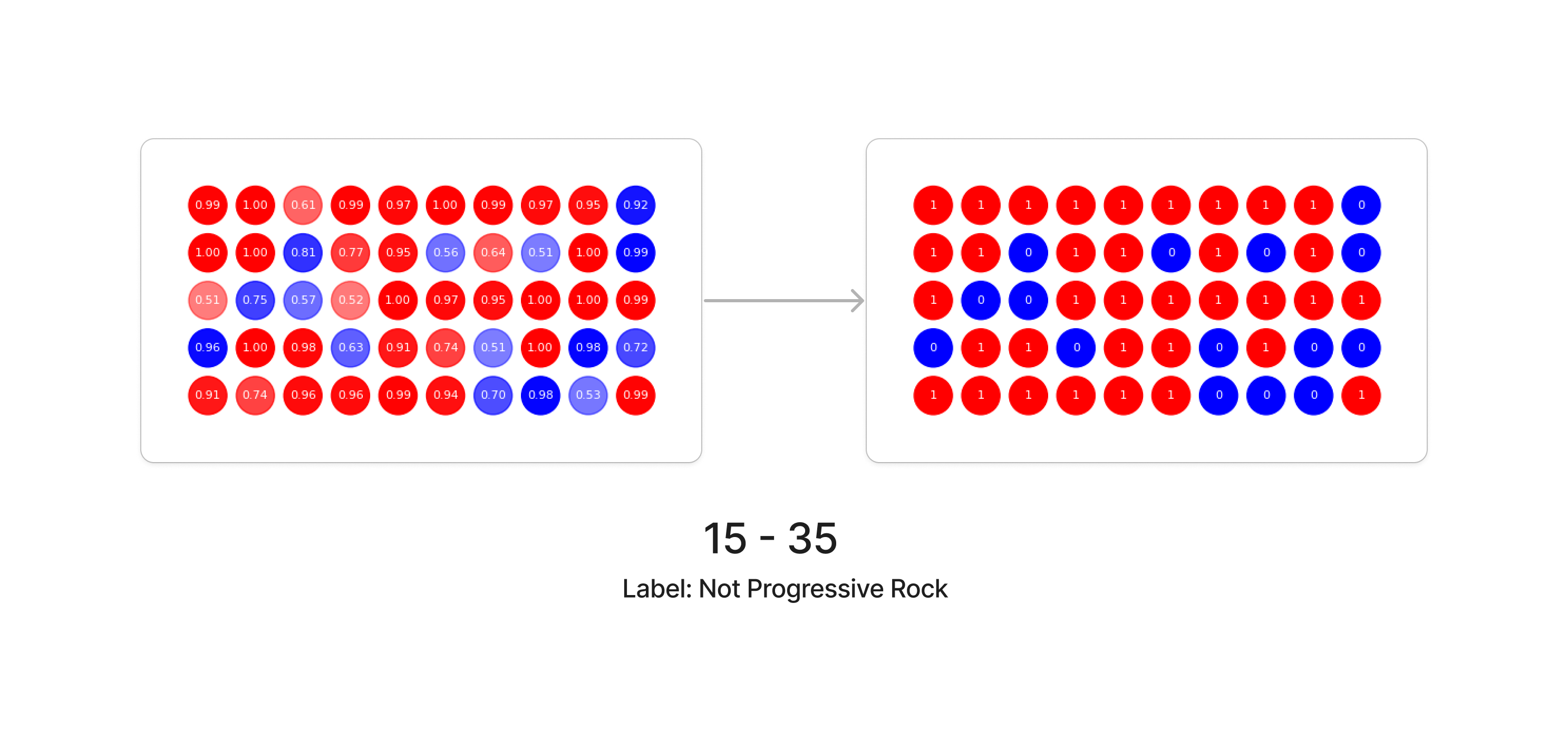}
    \caption{Classifying 50 Snippets of All the Stars by Kendrick Lamar \& SZA using our winner-take-all voting strategy}
    \label{nonprogsong}
\end{figure}

In this section, we describe a majority vote algorithm that was used to label the songs using the output of the classifier.
Each snippet of a song is run through our neural network, which assigns two probabilities of label \textit{prog} and \textit{non-prog}. 
If the probability of label \textit{prog} is greater than \textit{non-prog}, i.e if the classifier outputs a \textbf{prog} likelihood of greater than \textbf{0.5}, the snippet is classified as \textit{prog} and vice-versa for \textit{non-prog}.
Each classification result from the snippets contributes a vote towards the final song classification. A snippet classified as \textit{prog} adds a vote for \textit{prog}, and similarly, a \textit{non-prog} classification adds a vote for \textit{non-prog}.
After all, snippets are classified, and the votes are totaled. If the number of \textit{prog} votes exceeds the number of \textit{non-prog} votes, the song is classified as "Progressive Rock." (\ref{progsong})  Conversely, if \textit{non-prog} votes are in the majority, the song is classified as "Non Progressive Rock." (\ref{nonprogsong}) 
We

\chapter{Bagging, Boosting and Decision Trees}
\label{Chap:3}
\section{Ensemble Methods for Classification}

Using Ensemble Methods for Classification tasks is the approach of combining various models together with the goal of obtaining a system with improved accuracy and reduced over-fitting. The idea is that the combination of the multiple models is better than the output of the individual models themselves, and it works well with classification tasks that take input features and predict the outcome based on these features (ie. classifying \textit{prog} rock vs non prog rock given input features listed above). 

The two main categories that ensemble methods fall into are Bagging and Boosting. Bagging is the approach in which multiple models are trained using different subsets of the training set, with replacement, and every model performs classification and the majority vote decides on the final outcome. Some examples of Bagging techniques include Random Forest, Bagging, and ExtraTrees. Boosting, on the other hand, is the approach of training the various models in sequential order where the models learn from the mistakes of the previous training. This leads to a gradual increase in performance, and some examples of this approach include XGBoost and Gradient Boosting. All of these models mentioned can be used via the sklearn ensemble library.

In this project, we believed that using ensemble methods to approach the complex task at hand was worth exploring. We researched and trained multiple ensemble classification models with both boosting and bagging methods in mind, including Random Forest, Bagging, and ExtraTrees (bagging and decision trees) as well as XGBoost and Gradient Boosting (boosting). 

\section{PCA For Dimensionality Reduction}
Since the size and complexity of the dataset was large, before running some of the ensemble methods we needed to find a way to reduce the dimensionality of the dataset. We were conducting the project with limited computational resources and to tackle this problem we utilized Principle Component Analysis. By reducing the number of features, our models were more efficient and easier to run. A general trend that we saw was reducing the dataset to too few features (ie. 100) reduced the accuracy, so we settled for 200 features (despite the longer training time) for improved accuracy. 

\begin{figure}
    \centering
    \includegraphics[width=1\linewidth]{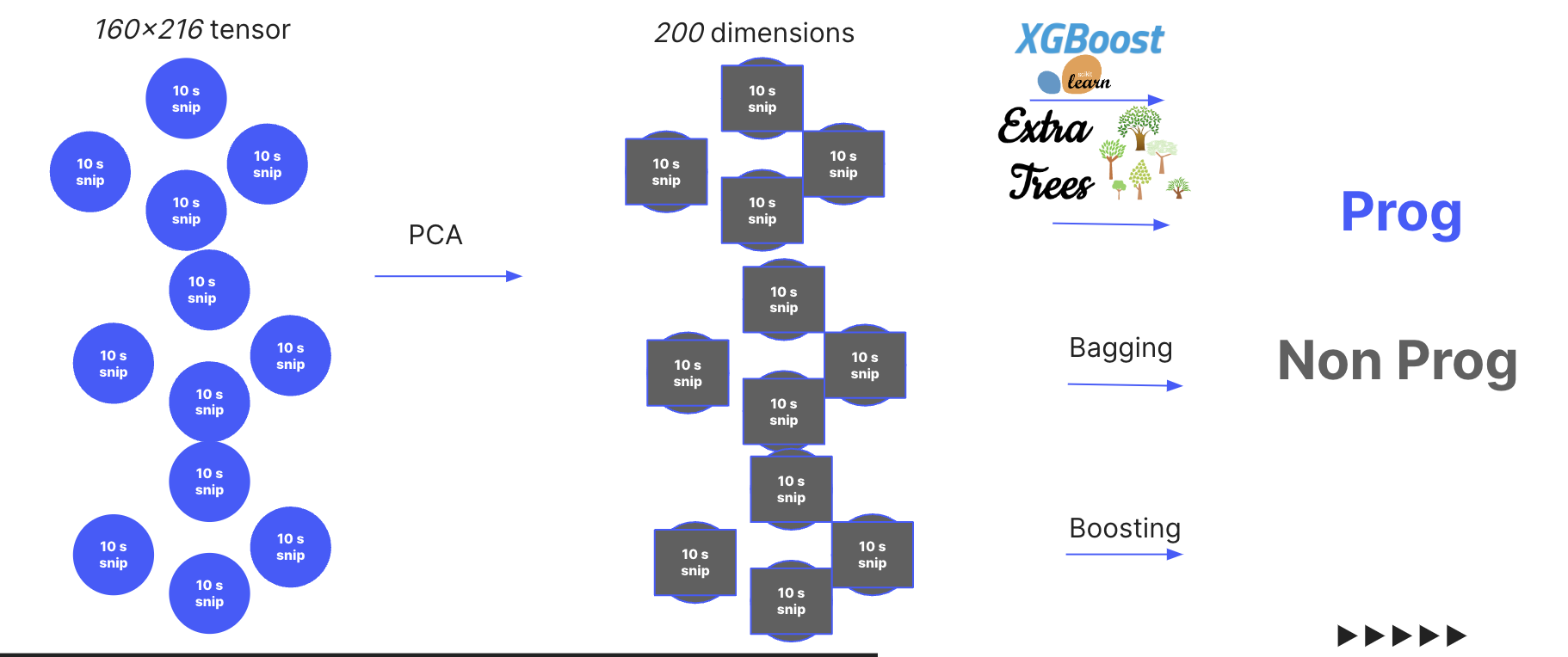}
    \caption{Flow chart describing how we've combined PCA with ensemble methods}
    \label{fig:pca-ensemble}
\end{figure}

\section{Bagging and Decision Trees (Random Forest, ExtraTrees, and Bagging)}
In this project, we implemented several bagging methods, where multiple models (decision trees in our case) were trained on subsets of the dataset and the output predictions were aggregated to improve model accuracy. 

The first method we employed was Random Forest for Classification. The reason for this choice was the ability of this classifier to handle complex datasets with many features, such as the features in our music classification project. At training time, multiple decision trees are created and the output of the model is the mean prediction of the classes. Since the Random Forest Classifier algorithms are generally fast to train, we decided not to use the PCA approach described for this bagging method, with the number of trees set in the forest set to 101. 

In similar fashion to Random Forest, we attempted to train an Extra Trees Model. This model builds many trees, but in contrast to Random Forest it selects the splits randomly rather than using the best split found in the subset. The purpose of using Extra Trees is adding this extra layer of randomness to the model and reducing the variance while doing so. Again, we set the number of different decision trees to 100 by setting the n estimators parameter to this value, and we avoided the use of PCA here due to the faster training nature of the model.  

Our last approach to explore bagging was with the use of the Bagging Classifier with the Decision Tree Classifier as a base estimator and the number of trees set to 50 due to limited computational resources. We also utilized PCA with the number of features equal to 100 due to extremely long training time. The way in which we executed this approach was the decision tree classifier was the individual model used repeatedly in the bagging ensemble method. 

During the training of these models, the training accuracy was nearly 100 percent for each method. All of these models were trained using 80 percent of the training dataset, as 20 percent was set aside for validation. The validation set approach as well as the test set approach will be explained later in the report.  

\begin{figure}
    \centering
    \includegraphics[width=0.75\linewidth]{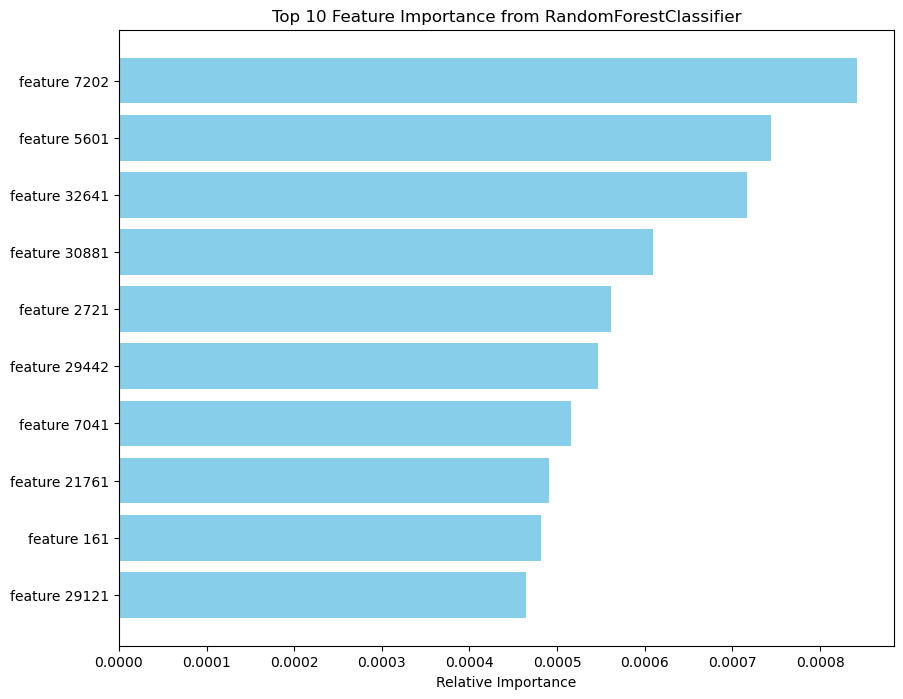}
    \caption{Random Forest feature importance out of 34560 features}
    \label{fig:enter-label}
\end{figure}

\section{Boosting (XGBoost and Gradient Boosting)}
We also implemented several boosting methods of ensemble learning including XGBoost and Gradient Boosting. Boosting is the approach in which models are built in sequence and builds off of the errors of each predecessor to improve the model accuracy while reducing bias. 

XGBoost, also known as eXtreme Gradient Boosting, is an enhanced and efficient implementation of gradient boosting. Known for its handling of large datasets, speed, and effectiveness in handling complex problems, we decided to run XGBoost with reduced data using PCA. The different runs we attempted using XGBoost include PCA with 100 and 200 features, as well as with 100 and 200 rounds of boosting. Generally, the model performed best with PCA to 200 features and 200 rounds of boosting, learning rate at 0.1 and max depth of 5 (due to limited computational resources). The training accuracy for this model ranged from 80-90 percent depending on the hyperparameters, and the validation and test results will be discussed in a later section. 

Lastly, we attempted the Gradient Boosting approach, a model designed for increased accuracy rather than the efficiency that XGBoost is known for. We used the Gradient Boosting Classifier from scikit-learn and explored training with n components of PCA ranging from 100-200 and the n estimators of Gradient Boosting also ranging from 100-200. We set the learning rate to 0.1, and the max depth to 3. The reason for the simplified hyper parameters here is due to gradient boostings decreased efficiency than what we saw in XGBoost. The training accuracy on the training set was about 77 percent, lower than before and the validation and test set results will be discussed with the results. 

\chapter{1D CNN Setup}
\label{Chap:4}
\section{Baseline}
1D Convolutional Neural Networks (1D CNNs) were developed initially to handle sequence data or any data with a temporal dimension. They are particularly suited for applications where the input data is naturally one-dimensional, such as audio signals, time series, or sequential data.
\cite{1dcnn}
1D CNNs are used in analysing biomedical signal like EEG or ECG signals and data from sensors, such as accelerometers or gyroscopes. \cite{kiranyaz20191d}
This gave us a good reason to explore 1D CNNs and code up the previous project as a baseline. 
We build a depthwise-CNN, starting off with 3 1D convolution layer with constant kernal size and padding. This is followed by a 3 Linear Layers. The model architecture consists of sequential convolutional layers that increase the number of channels from 160 to 256 and then to 512. Each convolutional layer uses a kernel size of 3, typical for capturing patterns in time-series data like audio. The stride of 1 and padding of 1 are used to maintain the dimensionality of the input through the layers. No pooling was utilized.

We employ the  \textbf{ReLU (Rectified Linear Unit)} as our activation function after each convolutional and fully connected layer except the last fully connected layer to introduce non-linearity into the model, helping it learn more complex patterns in the data. We use the \textbf{nn.CrossEntropyLoss()} as our loss function, implicitly a softmax to the target class probability. Adam \cite{kingma2017adam}, an adaptive learning rate optimization algorithm, is specifically engineered for the training of deep neural networks. It uniquely calculates individual learning rates for various parameters, enhancing the efficiency and effectiveness of the training process. We use a low learning rate of 0.001. 

We train the model over 10 epochs multiple times and end up with the training accuracy of about 72\% on the snippets on average. We evaluate the model on the validation set
and achieve an 55\% on the snippets. 
\begin{figure}
    \centering
    \includegraphics[width=1\linewidth]{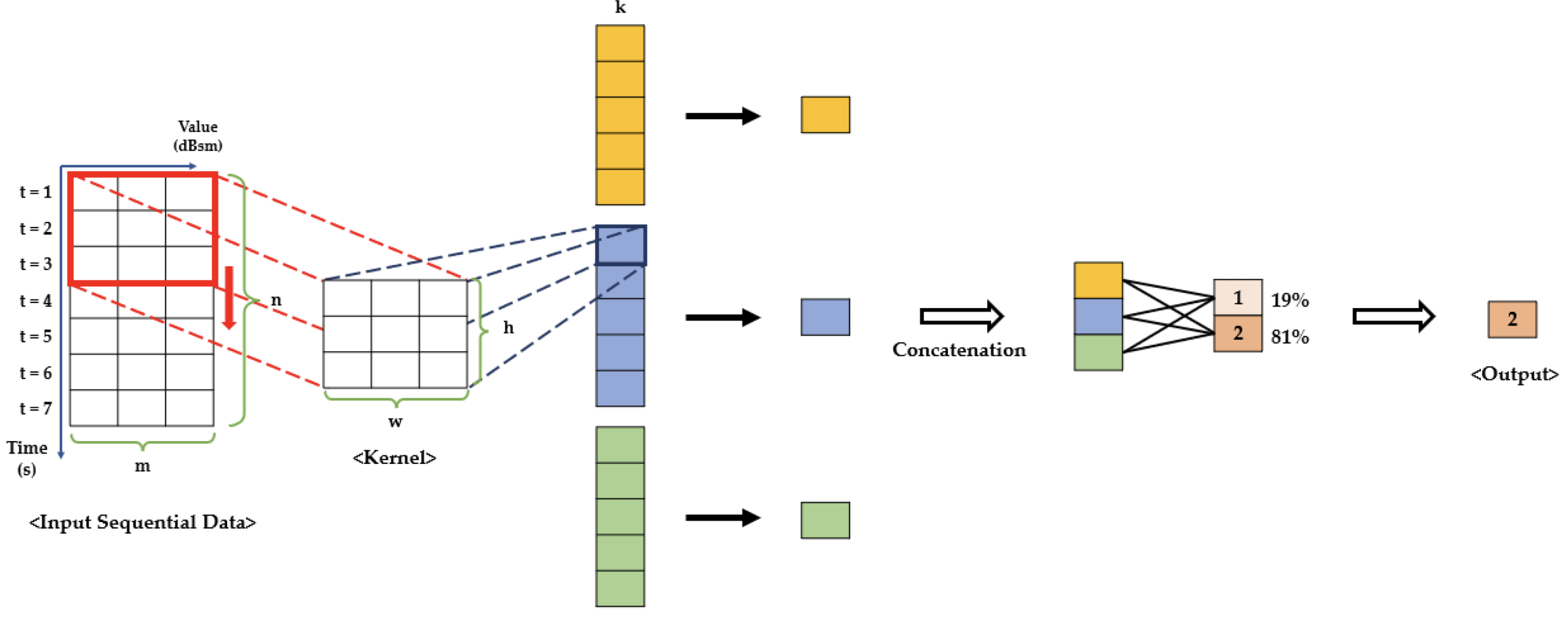}
    \caption{Diagram representing a 1D CNN model}
    \label{baseline}
\end{figure}

\chapter{1D CNNs Model Architecture}
\label{Chap:5}
Here, we present the architecture of our two best performing 1D deep convolutional neural networks. We have decided to name these two networks as \textit{Zuck} and \textit{Satya} to honor our two modern overlords. Standing on the shoulders of the giants that were previous CAP 6610 students, we decided to stick with the \textit{Adam} optimizer with a learning rate of $10^{-3}$ and an epoch size of 10 for both networks.

\section{\textit{Zuck}}

\begin{figure}
    \centering
    \includegraphics[width=0.9\linewidth]{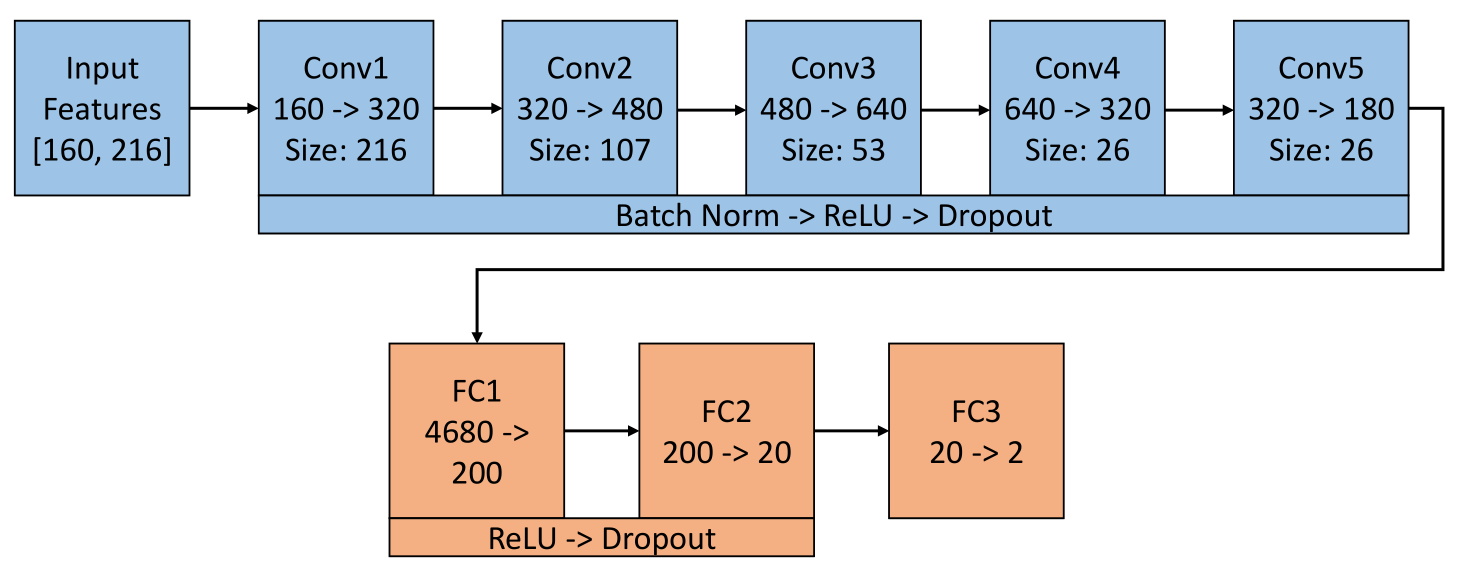}
    \caption{Simple architecture of the \textit{Zuck} model}
    \label{cnn-simple}
\end{figure}

\begin{figure}
    \centering
    \includegraphics[width=\linewidth]{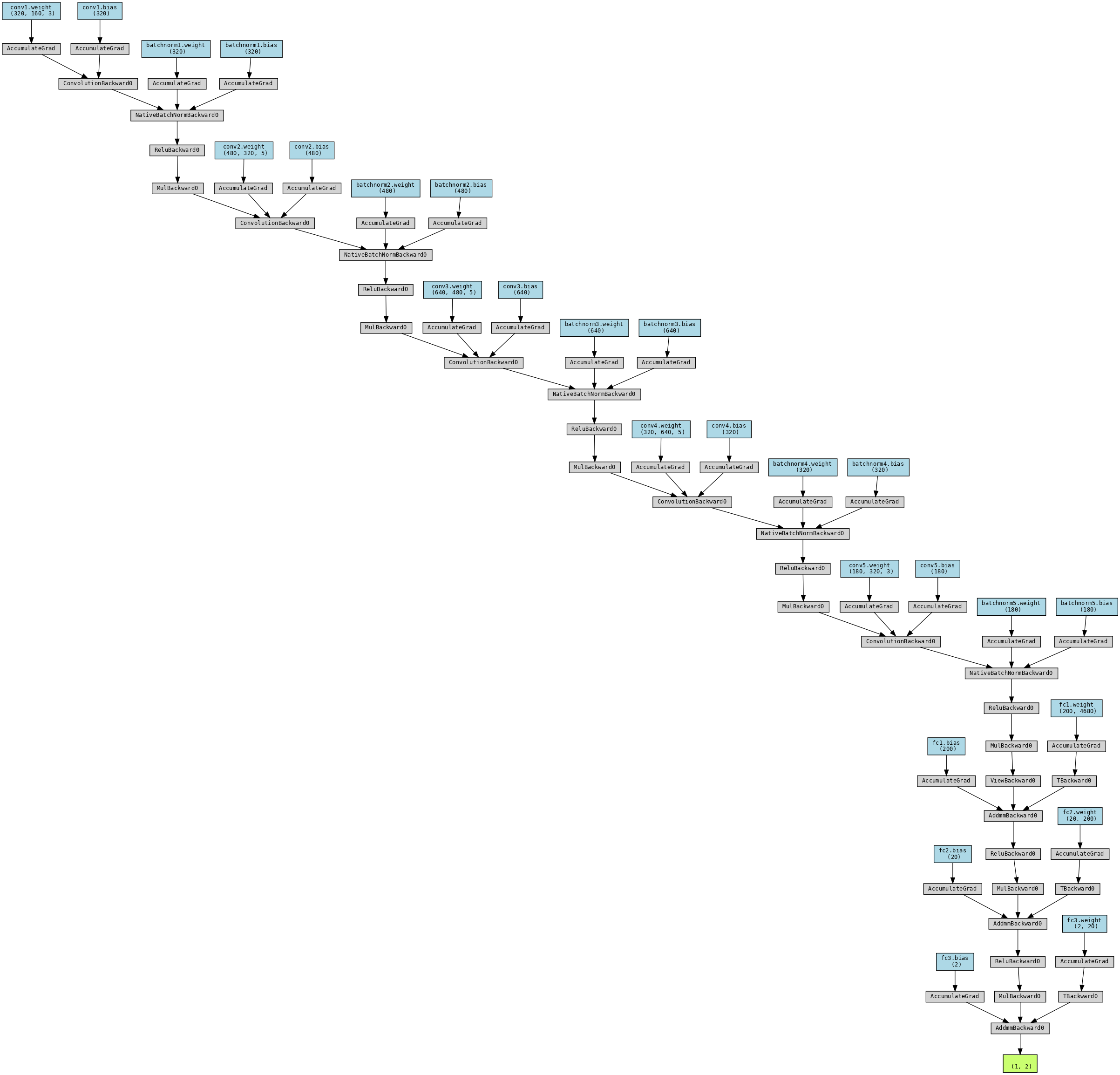}
    \caption{Full architecture of the \textit{Zuck} model visualized using \textit{torchviz}.}
    \label{cnn-full}
\end{figure}

The architecture of the \textit{Zuck} 1D deep convolutional neural network is presented in Figures \ref{cnn-simple} and \ref{cnn-full}. The network consists of five convolutional layers, each followed by batch normalization and ReLU activation. Following these layers, the output is flattened into a 1D vector, serving as input to three fully connected layers: two hidden layers with ReLU activation and one output layer. We used our winner-take-all voting strategy to predict the genre of the song based on the classification of each snippet in the output layer.

The input layer is a tensor of shape of [batch\_size, in\_channels, seq\_length]. The input data shape of each sample is [160, 216] where 216 is the sequence length and 160 is the number of input channels. These are then fed to five convolutional layers with varying kernel sizes, and strides. The first convolutional layer with output channels of 320 and kernel size of 3 preserves the input size of 216. Then, the second, third and fourth convolutional layers, with a stride of 2 and a kernel size of 5 progressively reduce the dimensions to 107, 53, and 26, respectively. Finally, the last convolutional layer maintains the size at 26. The feature maps are then merged to a single vector of size 4680, which was then fed to the 3 fully connected hidden layers.

For non-linearity, each convolutional layer and the first two hidden layers were followed by a ReLU activation function. Batch normalization is implemented after each convolution layer to normalize and stabilize the activations. Similarly, a dropout technique with a probability of 0.25 was used after each activation to prevent overfitting. 

\begin{figure}
    \centering
    \includegraphics[width=0.75\linewidth]{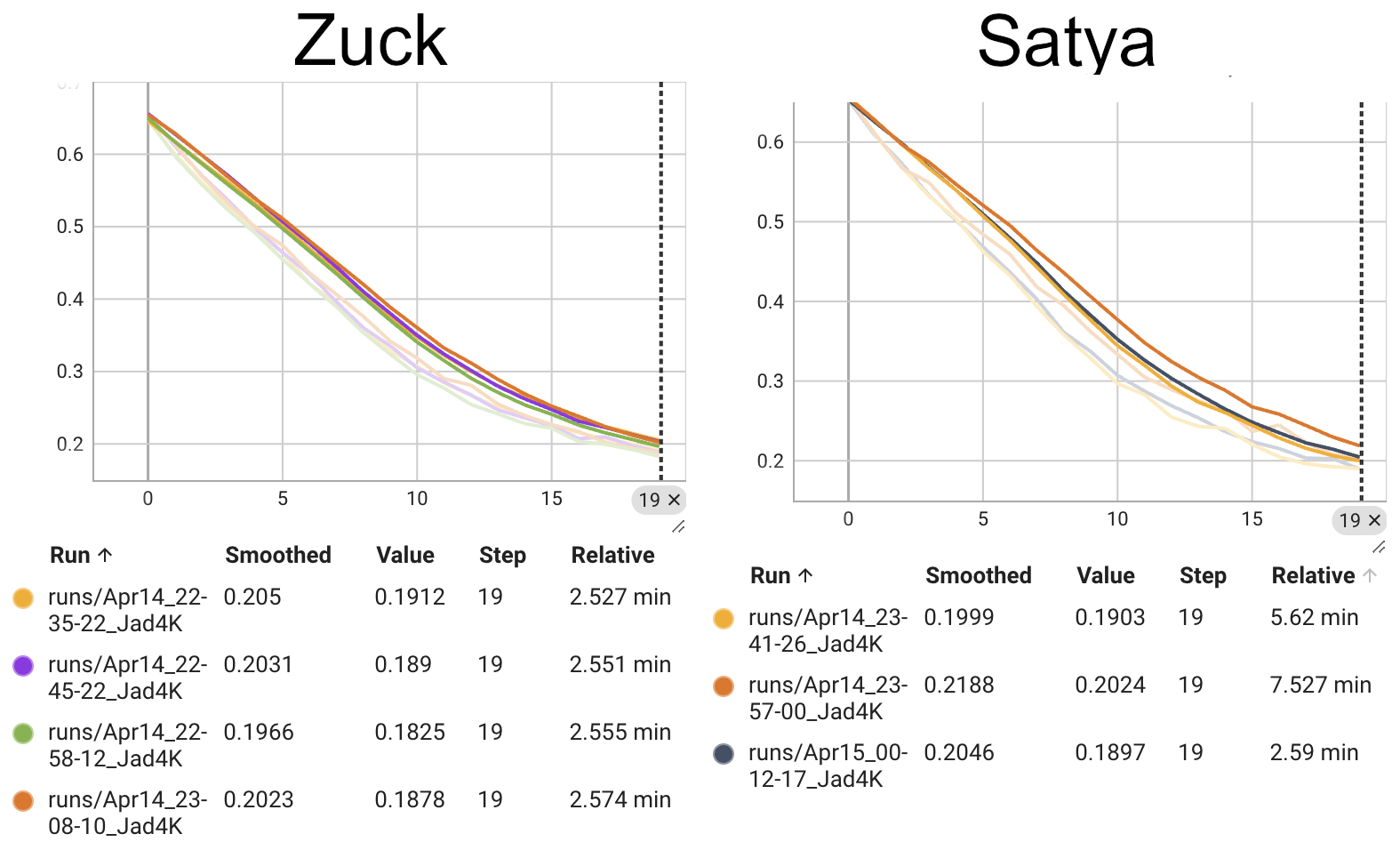}
    \caption{Models \textit{Zuck} and \textit{Satya} loss plots over 3 training runs}
    \label{fig:zuck-loss}
\end{figure}

\section{\textit{Satya}}

The \textit{Satya} model is structured similarly to the \textit{Zuck} model. It is composed of five convolutional layers, each complemented by an activation function to ensure effective non-linearity and a dropout of 0.2 to prevent overfitting. Starting with a sequence length of 216, the network reduces the dimensions to 52 before flattening to a vector of size 6784 after the fifth convolutional layer. These dilated convolutions aim to capture larger context without significantly increasing the computational cost.

\begin{figure}
    \centering
    \includegraphics[width=0.9\linewidth]{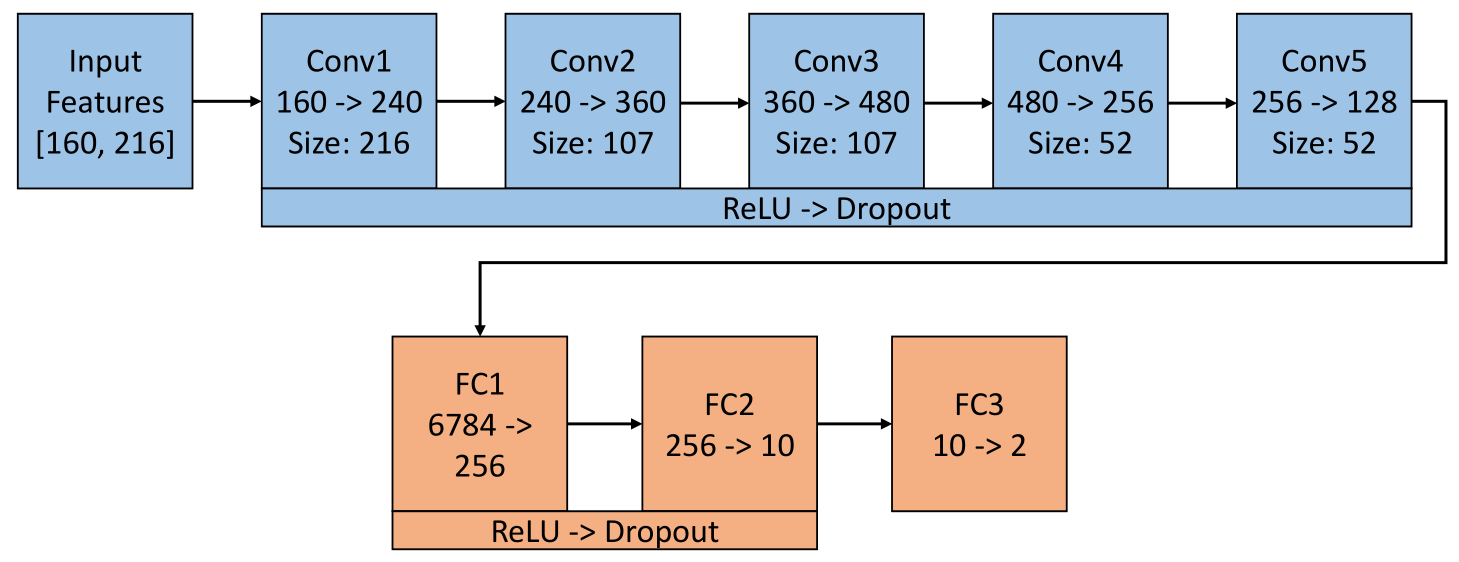}
    \caption{Simple architecture of the \textit{Satya} model}
    \label{cnn-satya}
\end{figure}

The key difference between the two models lies in the learning rate used for \textit{Satya} model. Here, we employ a learning rate scheduler unlike the \textit{Zuck} model which used a single value. Specifically, we combined a linear warm-up phase, where we initially allowed the learning rate to increase gradually from zero to a set value of $10^{-3}$, with a cosine decay to adjust the learning rate over epochs. This approach aims to have a stable training during early epochs, and further fine-tuning using progressively smaller updates to prevent overshooting.

\begin{figure}
    \centering
    \includegraphics[width=0.7\linewidth]{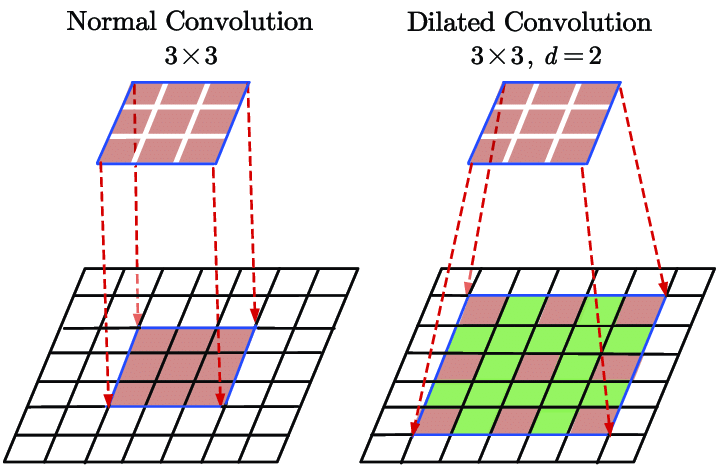}
    \caption{Dilated Convolution}
    \label{fig:}
\end{figure}

Also, dilated convolution is strategically implemented in one of the middle layers. This placement is crucial as it allows the network to capture broader contextual information in the intermediate stages of processing, which is essential for tasks requiring detailed spatial understanding.

\chapter{Audio Spectrogram Transformer}
\label{Chap:6}
CNNs have been the core framework for audio classification for the past decade. However, in recent times, the use of the Transformer architecture, which has self-attention \cite{vaswani2023attention} at its heart, has become widespread. The Audio Spectrogram
Transformer (AST), a convolution-free, purely attention-based
model \cite{gong2021ast} for audio classification caught our eye as it was the first convolution-free and purely attention based model which featured a simple architecture and superior SOTA performance on audio classification datasets. It is closely related to the ViT Transformer architecture \cite{dosovitskiy2021image} for vision tasks. A difference being that the ViT has only been applied to fixed-dimensional inputs (images), while AST can process variable-length audio inputs. 

We now discuss the specifics of the working of the model.
First, the raw input audio waveform of 5
seconds is converted into a sequence of 128-dimensional log
Mel-filterbank. This results in a 128×128 spectrogram. We then split the spectrogram into a
sequence of 16×16 patches with an overlap of 6 in both time and frequency dimension. These 16x16 patches are then linearly projected into 1-D patch embeddings of size 768 to be read by the transformer encoder. In order to maintain the spatial structure of the 2D audio spectrogram, it's important for the model to understand the sequence or order of the patches. However, since transformers don't inherently capture this information, positional embeddings are added to the patch embeddings. These embeddings are of the same size (768), which helps the model effectively comprehend the order of the input data.
Unlike traditional models that may use convolutional layers for initial feature extraction, AST employs a transformer architecture that relies solely on self-attention to process the input embeddings. This allows the model to capture more complex patterns and dependencies in the audio data.

\begin{figure}
    \centering
    \includegraphics[width=0.45\linewidth]{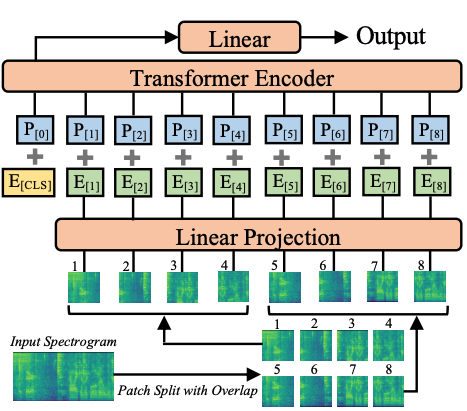}
    \caption{AST: Audio Spectrogram Transformer
AST, by MIT Computer Science and Artificial Intelligence Laboratory
2021 InterSpeech}
    \label{fig:enter-label}
\end{figure}

For our implementation, we use an AST model that was finetuned on AudioSet, \cite{7952261} 632 audio event classes and a collection of 2,084,320 human-labeled 10-second sound clips drawn from YouTube. A \href{https://huggingface.co/MIT/ast-finetuned-audioset-10-10-0.4593}{Hugging Face} implementation is downloaded and finetuned on the dataset. We train the model and save over 10 checkpoints. 

Periodically, the model's performance is evaluated on a validation set to monitor its generalization to unseen data and to prevent overfitting. We use a very small learning rate of $3 \times 10^{-5}$.

\chapter{Results}
\label{Chap:7}
\section{Ensemble Methods}
\subsection{Validation}
Initially, we divided the dataset into a training and validation set with an 80/20 split. Below are the results for the validation set, which were run after training the models, using the winner-take-all voting strategy described earlier. 

The validation set performance for the boosting methods are presented in Table \ref{table:ensemble_val_boosting}, and the validation set performance for the bagging methods are presented in Table \ref{table:ensemble_val_bagging}

\begin{table}[h]
\centering
\caption{Boosting Performance Metrics and Confusion Matrices For Validation Set}
\begin{tabular}{|c|c|c|c|c|}
\hline
\textbf{Model} & \textbf{Accuracy} & \textbf{Precision} & \textbf{Recall} & \textbf{Confusion Matrix} \\
\hline
XGBoost + PCA & 93.26\% & 93.33\% & 87.5\% & \begin{tabular}[c]{@{}c@{}}Prog True: 28\\ Prog False: 2\\ Non-Prog True: 55\\ Non-Prog False: 4\end{tabular} \\
\hline
Gradient Boosting + PCA & 82.02\% & 93.33\% & 66.66\% & \begin{tabular}[c]{@{}c@{}}Prog True: 28\\ Prog False: 2\\ Non-Prog True: 45\\ Non-Prog False: 14\end{tabular} \\
\hline
\end{tabular}
\label{table:ensemble_val_boosting}
\end{table}

\begin{table}[h]
\centering
\caption{Bagging Performance Metrics and Confusion Matrices For Validation Set}
\begin{tabular}{|c|c|c|c|c|}
\hline
\textbf{Model} & \textbf{Accuracy} & \textbf{Precision} & \textbf{Recall} & \textbf{Confusion Matrix} \\
\hline
Random Forest & 95.51\% & 93.33\% & 93.33\% & \begin{tabular}[c]{@{}c@{}}Prog True: 28\\ Prog False: 2\\ Non-Prog True: 57\\ Non-Prog False: 2\end{tabular} \\
\hline
ExtraTrees & 94.38\% & 90.00\% & 93.10\% & \begin{tabular}[c]{@{}c@{}}Prog True: 27\\ Prog False: 3\\ Non-Prog True: 57\\ Non-Prog False: 2\end{tabular} \\
\hline
\hline
Bagging + PCA & 94.38\% & 93.33\% & 90.32\% & \begin{tabular}[c]{@{}c@{}}Prog True: 28\\ Prog False: 2\\ Non-Prog True: 56\\ Non-Prog False: 3\end{tabular} \\
\hline
\end{tabular}
\label{table:ensemble_val_bagging}
\end{table}

\subsection{Test}
Utilizing the test set tensors, which were curated in the same way as the validation tensors, we classified the songs as prog/non-prog using the winner take all voting strategy described previously. 

The test set performance for the boosting methods are presented in Table \ref{table:ensemble_test_boosting}, and the test set performance for the bagging methods are presented in Table \ref{table:ensemble_test_bagging}

\begin{table}[h]
\centering
\caption{Boosting Performance Metrics and Confusion Matrices For Test Set}
\begin{tabular}{|c|c|c|c|c|}
\hline
\textbf{Model} & \textbf{Accuracy} & \textbf{Precision} & \textbf{Recall} & \textbf{Confusion Matrix} \\
\hline
XGBoost + PCA & 71.57\% & 78.99\% & 69.43\% & \begin{tabular}[c]{@{}c@{}}Prog True: 109\\ Prog False: 29\\ Non-Prog True: 85\\ Non-Prog False: 48\end{tabular} \\
\hline
Gradient Boosting + PCA & 70.48\% & 77.54\% & 68.59\% & \begin{tabular}[c]{@{}c@{}}Prog True: 107\\ Prog False: 31\\ Non-Prog True: 84\\ Non-Prog False: 49\end{tabular} \\
\hline
\end{tabular}
\label{table:ensemble_test_boosting}
\end{table}

\begin{table}[h]
\centering
\caption{Bagging Performance Metrics and Confusion Matrices For Test Set}
\begin{tabular}{|c|c|c|c|c|}
\hline
\textbf{Model} & \textbf{Accuracy} & \textbf{Precision} & \textbf{Recall} & \textbf{Confusion Matrix} \\
\hline
Random Forest & 74.54\% & 71.01\% & 77.17\% & \begin{tabular}[c]{@{}c@{}}Prog True: 98\\ Prog False: 40\\ Non-Prog True: 29\\ Non-Prog False: 104\end{tabular} \\
\hline
ExtraTrees & 76.38\% & 73.19\% &78.91\% & \begin{tabular}[c]{@{}c@{}}Prog True: 101\\ Prog False: 37\\ Non-Prog True: 106\\ Non-Prog False: 27\end{tabular} \\
\hline
\hline
Bagging + PCA & 67.89\% & 67.39\% & 68.89\% & \begin{tabular}[c]{@{}c@{}}Prog True: 93\\ Prog False: 45\\ Non-Prog True: 91\\ Non-Prog False: 42\end{tabular} \\
\hline
\end{tabular}
\label{table:ensemble_test_bagging}
\end{table}

\section{1D CNNs}

\subsection{Validation}

To assess the performance of our models, we initially divided the dataset into a training set (80\%) and a validation set (20\%). Here, we record the results of various 1D CNN models against the validation set which served as the base models for our \textit{Zuck} and \textit{Satya} 1D CNNs. 

\begin{table}[!h]
\caption[1D CNN for Validation]{The top 4 performing 1D CNNs for the 89 songs in the validation set.}
\centering
\begin{tabular}{|c|c|c|c|}
\hline
\textbf{Model}     & \textbf{Conv Layer}                                                                                                                                                                  & \textbf{Hidden Layer}                                                  & \textbf{Activation Func.}   \\ \hline
\multirow{2}{*}{1} & \multirow{2}{*}{160 -\textgreater 32 -\textgreater 64 -\textgreater 128 -\textgreater 256 -\textgreater 128 -\textgreater 64}                                                        & \multirow{2}{*}{100 -\textgreater 10 -\textgreater 2}                  & \multirow{2}{*}{ReLu}       \\
                   &                                                                                                                                                                                      &                                                                        &                             \\ \hline
\multirow{2}{*}{2} & \multirow{2}{*}{160 -\textgreater 32 -\textgreater 64 -\textgreater 128 -\textgreater 256 -\textgreater 128 -\textgreater 64}                                                        & \multirow{2}{*}{100 -\textgreater 50 -\textgreater 10 -\textgreater 2} & \multirow{2}{*}{ReLu}       \\
                   &                                                                                                                                                                                      &                                                                        &                             \\ \hline
\multirow{2}{*}{3} & \multirow{2}{*}{\begin{tabular}[c]{@{}c@{}}160 -\textgreater 32 -\textgreater 64 -\textgreater 128 -\textgreater 256 -\textgreater 128 -\textgreater 64\\ (Normalized)\end{tabular}} & \multirow{2}{*}{100 -\textgreater 10 -\textgreater 2}                  & \multirow{2}{*}{ReLu}       \\
                   &                                                                                                                                                                                      &                                                                        &                             \\ \hline
\multirow{2}{*}{4} & \multirow{2}{*}{160 -\textgreater 32 -\textgreater 64 -\textgreater 128 -\textgreater 256 -\textgreater 128 -\textgreater 64}                                                        & \multirow{2}{*}{100 -\textgreater 10 -\textgreater 2}                  & \multirow{2}{*}{leaky ReLu} \\
                   &                                                                                                                                                                                      &                                                                        &                             \\ \hline
\end{tabular}
\label{table:1dcnn_valid}
\end{table}

Presented in Table \ref{table:1dcnn_valid} are the top performing models on 89 songs in the validation set. Each model uses the \textit{Adam} optimizer with a learning rate of $10^{-3}$ and an epoch size of 10. The models also employ similar convolutional arithmetic with no reduction in dimensions, with a $216 \times 64$ size for the flattened vector after the final convolutional layer. This test aims to know the effectiveness of different activation functions, and to introduce normalization for more stable activation. We also added variations in the hidden layers to explore different complexities during training.

\begin{figure}[!h]
  \centering\noindent
  \includegraphics[width=0.75\textwidth,clip]{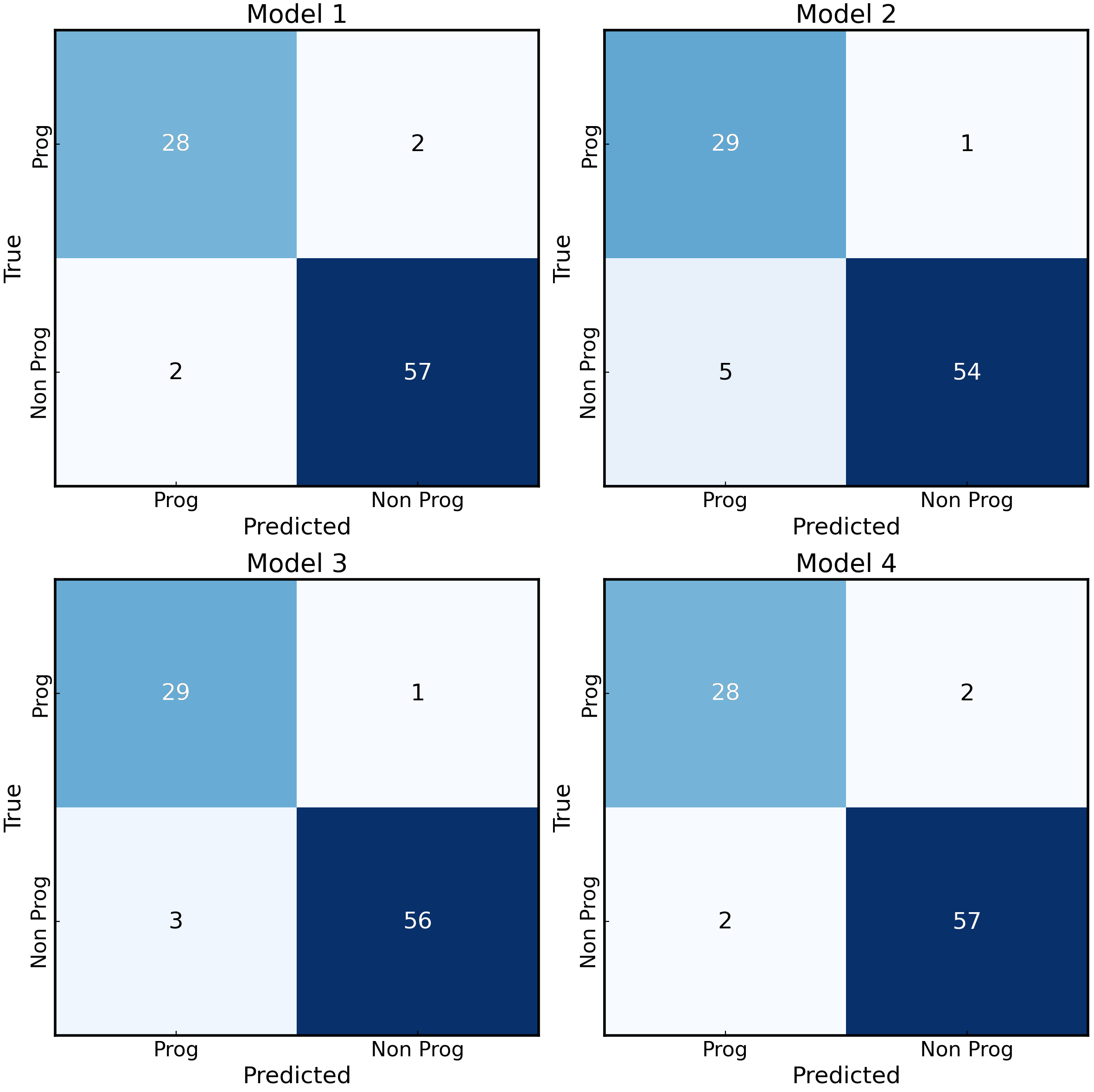}
  \caption[1D CNN Confusion Matrix for Validation]{Confusion matrices for the top 4 performing models on the validation set}
  \label{fig:confusion_validation}
\end{figure}

Applying these models to the validation set show that Models 1 and 4 show the best balance between sensitivity and specificity. The primary distinction between them lies in their activation functions, indicating how the choice between ReLu and LeakyReLU does not have a major effect on the model's performance. As such, the final models we built for the test set only used the ReLU activation function. Model 3 also follows closely, with a slightly better focus on sensitivity. This observation led us to implement batch normalization across all convolutional layers and the first two hidden layers in our \textit{Zuck} network. On the other hand, Model 2, while still effective, suggests the benefits of continuing with a three-hidden-layer architecture. Overall, these results provide us a strategy in designing and refining our final models for the test set.

\subsection{\textit{Zuck} and \textit{Satya}}

\begin{figure}[!h]
  \centering\noindent
  \includegraphics[width=0.75\textwidth,clip]{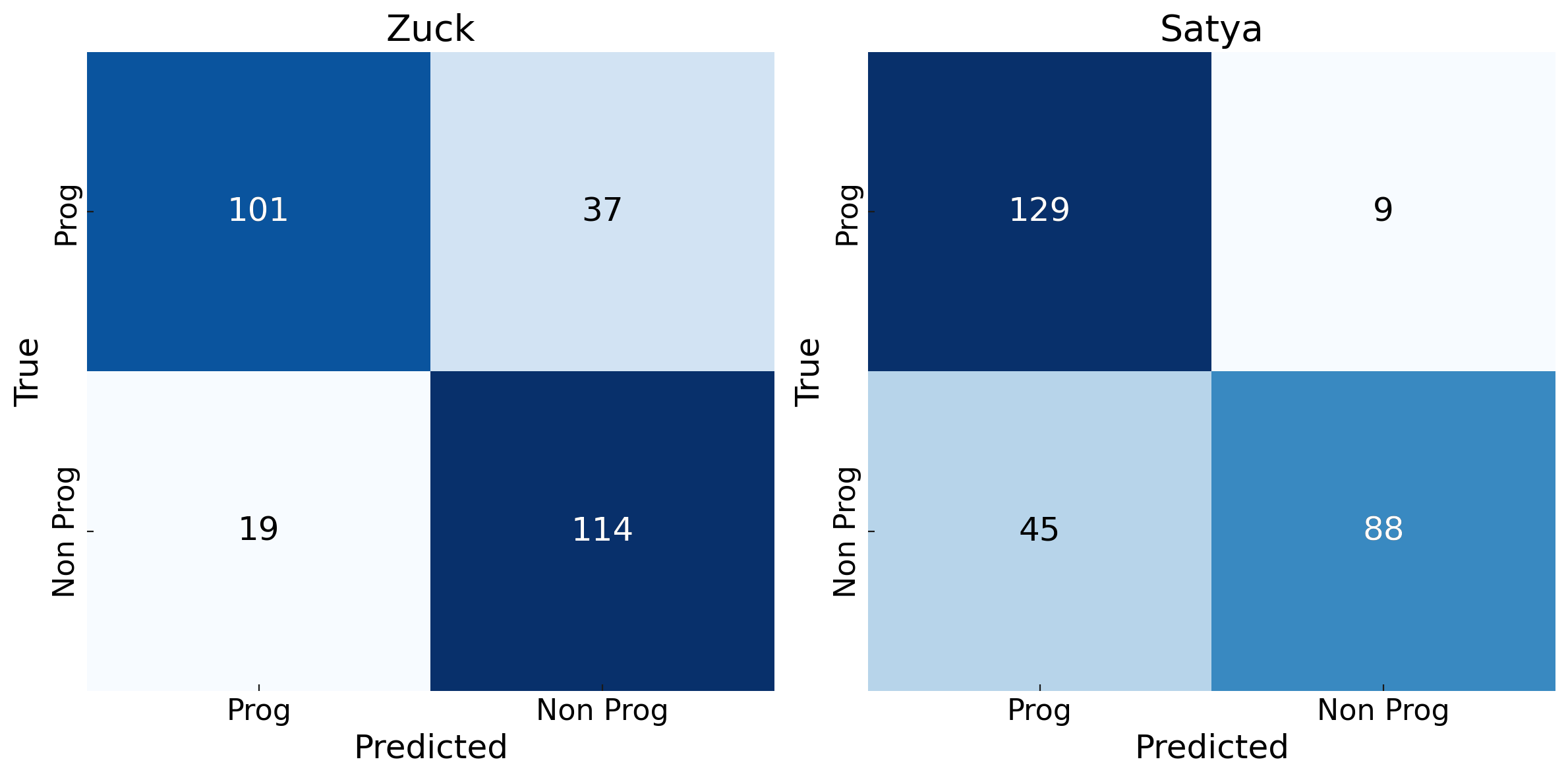}
  \caption[Zuck and Satya 1D CNN]{Confusion matrices for the performance of the \textit{Zuck} and \textit{Satya} models on the test set}
  \label{fig:zuck_satya_test}
\end{figure}

The confusion matrices for \textit{Zuck} and \textit{Satya} provide insights into the way these two models ended up being trained (Figure \ref{fig:zuck_satya_test}). The \textit{Zuck} 1D CNN seems to demonstrate a more conservative behavior, correctly identifying 114 non-prog and 101 prog rock songs while providing a high precision value of \textbf{84.2 \%}. This value comes at the expense of overlooking some prog rock songs while prioritizing rejection of non-prog tracks.

The \textit{Satya} model, on the other hand, is more liberal (Table \ref{table:satya_test}). The model seems to adapt an aggressive strategy in identifying prog rock songs, thus giving a high accuracy (\textbf{80.07 \%}) in place of a low precision score (\textbf{74.14 \%}) or a higher false negative rate.  It mislabeled only 9 prog songs as non-prog, indicating that this model has been trained to minimize the misses in the prog genre.

These results suggest that the quality of these models hinges on their particular use. The \textit{Zuck} model is more cautious in its prediction, thus is better at minimizing false positives or wrong predictions regarding non-prog songs. Likewise, \textit{Satya} is more applicable if we want to pull more prog rock songs from a diverse set of music genres. Given these results, we further hypothesize that although \textit{Satya} would be better at discriminating prog rock from a mixed genre set, \textit{Zuck} would excel more in specific applications where the test set specifically contains non-prog rocks that stemmed directly from prog rock (e.g. songs of progressive pop, post rock, math rock genres).

\begin{table}[h!]
\centering

\label{tab:performance_metrics}
\begin{tabular}{lcc}
\toprule
\textbf{Metric} & \textbf{Snippets} & \textbf{Songs} \\
\midrule
Accuracy & 65.41\% - 67.87\%  & 75.28\% - \textbf{80.07\%} \\
Precision (Prog) & 73.67\% - 75.01\% & 89.85\% - 93.47\% \\
Recall (Prog) & 71.81\% - 74.01\% &  70.05\% - 74.13\% \\
\midrule
\multicolumn{3}{c}{\textbf{Confusion Matrix (Snippets)}} \\
\midrule
& Predicted Prog & Predicted Non Prog \\
Actual Prog & 5037 & 1678 \\
Actual Non Prog & 1768 & 2244 \\
\midrule
\multicolumn{3}{c}{\textbf{Confusion Matrix (Songs)}} \\
\midrule
& Predicted Prog & Predicted Non Prog \\
Actual Prog & 129 & 9 \\
Actual Non Prog & 45 & 88 \\
\bottomrule
\end{tabular}
\caption{Performance Metrics and Confusion Matrices for Satya on Test Set including confusion matrix for best run}
\label{table:satya_test}
\end{table}

\section{Audio Spectrogram Transformer}

We tried our fine-tuned model on the test set. We ended up with the following results. 
The model demonstrates a strong ability to classify \textit{Prog} snippets with high recall, indicating it can identify most of the positive class instances However, the precision is slightly lower, suggesting a moderate number of false positives (as we've seen in the previous years of this project) . The overall accuracy is close to 79.4\%, which is a solid performance.

What makes this model a good classifier that it classifies \textit{non-prog} snippets most accurately. 81.25\% of the non-prog were classified correctly, most by any model so far. 

Misclassified songs include: Little Man (most misclassified song), Peace To The Mountain by Coheed and Cambria, The Temple On the Edge of Time.

On the songs, the model shows a higher accuracy of approximately \textbf{85.4\%} compared to the Snippets. The precision is notably high, indicating that when the model predicts a song as \textit{Prog}, it is very likely to be correct. The recall is also strong, though there is some room for improvement in reducing false negatives. The F1 score is high, suggesting a balanced performance between precision and recall.

\begin{table}[h!]
\centering

\label{tab:performance_metrics}
\begin{tabular}{lcc}
\toprule
\textbf{Metric} & \textbf{Snippets} & \textbf{Songs} \\
\midrule
Accuracy & 77.46\% & 85.87\% \\
Precision (Prog) & 79.28\% & 90.29\% \\
Recall (Prog) & 83.76\% & 83.44\% \\
\midrule
\multicolumn{3}{c}{\textbf{Confusion Matrix (Snippets)}} \\
\midrule
& Predicted Prog & Predicted Non Prog \\
Actual Prog & 11402 & 2979 \\
Actual Non Prog & 2210 & 6438 \\
\midrule
\multicolumn{3}{c}{\textbf{Confusion Matrix (Songs)}} \\
\midrule
& Predicted Prog & Predicted Non Prog \\
Actual Prog & 121 & 13 \\
Actual Non Prog & 24 & 104 \\
\bottomrule
\end{tabular}
\caption{Performance Metrics and Confusion Matrices for Test Sets}
\end{table}

\begin{figure}
    \centering
    \includegraphics[width=0.0\linewidth]{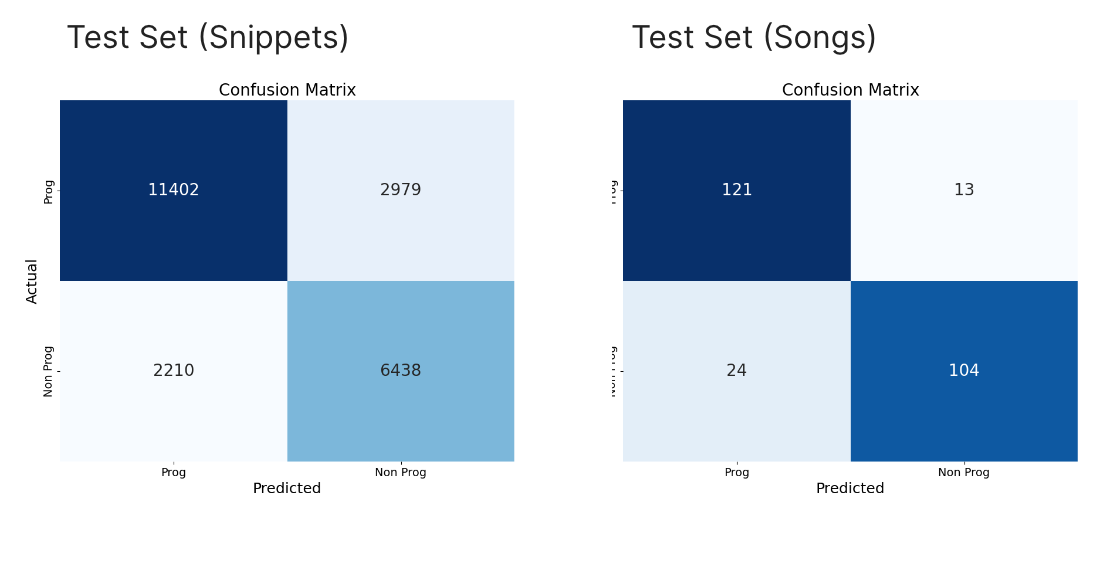}
    \caption{Confusion matrix of the AST on the test set}
    \label{fig:ast-results}
\end{figure}

\section{Post Prog}

Here we discuss the contents of Table 7.3, which is a list of post-prog songs along with the label assigned. We classify merely 4 out of 22 songs as non-prog while ideally all of these songs are non-prog. This is not surprising given that these songs are of mathcore, djent, grindcore, and post-prog categories which stemmed directly from the prog rock genre. This is also a possible consequence of the biggest assumption we had, which is that music classification properties supervene on audio properties. These labels, especially the non-prog labels on post-prog rock songs, can be arbitrary due to the historical and stylistic connection between any musical genre and its corresponding \textit{post} genre. The band Radiohead, for example, is infamous for answering, "No. We all \textit{hate} progressive rock music.", when asked if their album \textit{OK Computer} was influenced by the works of progressive rock bands such as Genesis and Pink Floyd. \cite{Sanneh_2017} In this case, history weighs heavier than the stylish connections between their songs and the prog rock genre.

\begin{table}[h]
\centering
\begin{tabular}{ll}
\hline
\textbf{Name} & \textbf{Label} \\
\hline
03 - Language II Conspire\_37.wav & prog \\
AFTER THE BURIAL - A Wolf Amongst Ravens\_32.wav & prog \\
05. Physical Education\_54.wav & prog \\
MONUMENTS - I The Creator\_06.wav & prog \\
CHIMP SPANNER - Bad Code\_40.wav & prog \\
01 Arithmophobia\_13.wav & prog \\
07 The Race Is About To Begin\_04.wav & prog \\
Meshuggah- Soul Burn\_13.wav & prog \\
06 A Light Will Shine\_02.wav & prog \\
PERIPHERY - Zyglrox\_02.wav & prog \\
Cloudkicker - Let yourself be huge\_06.wav & prog \\
Veil Of Maya - Punisher\_09.wav & prog \\
06-1289 Voyeur Will Shine Fight For Distinction Evolution Is Mine.\_60.wav & prog \\
Textures - Laments Of An Icarus\_49.wav & prog \\
Hacktivist - DECEIVE AND DEFY\_17.wav & prog \\
SikTh - Hold My Finger\_61.wav & prog \\
08-the\_haarp\_machine-machine\_over\_44.wav & prog \\
b1-master\_boot\_record-dma\_4\_cascade\_10.wav & prog \\
HEART OF A COWARD - Hollow\_05.wav & non-prog \\
01-darko\_us-splinter\_cell\_45.wav & non-prog \\
The Algorithm - Isometry\_25.wav & non-prog \\
BORN OF OSIRIS - Divergency\_44.wav & non-prog \\
\hline
\end{tabular}
\caption{Music Tracks and their Predicted Labels}
\label{table:music_tracks}
\end{table}

\chapter{Conclusion}

The problem of Music Genre classification is a difficult one with notable work done in the field of supervised learning to make machines differentiate between our music categories. 

Our pre-processing step involved extracting features like spectrograms, MFCCs, chromagrams, and beat-on-set positions from raw audio, and these were used as input to various classifiers. However, it is to be noted that these may not be the best features to work with, and work can be done to investigate which features would best describe music.

We choose to work with 1D CNNs given the time-series nature of audio data. However, there is no reason to claim that 1D CNNs perform better than 2D CNNs on audio data. Future work should try comparing similar 2D CNN models compared to 1D CNN baselines. 

Transformer architecture has become increasingly popular in recent years. It is critical that future research focus on transformers and their ability to handle large context. We attempted to collaborate with CNNs and Transformers but were unable to produce tangible results.

Unsupervised and semi-supervised learning methods should also be experimented with, given that they are objectively better than supervised methods and require far less data. Given an opportunity, we would focus on semi-supervised methods like Noisy student training for music tagging, transfer learning, or different pre-trained models.

We leave these findings and lessons to future CAP6610 students.

\printbibliography

\end{document}